# Solutions of the spherically symmetric SU(2) Einstein–Yang–Mills equations defined in the far field

Arthur G. Wasserman

*Department of Mathematics, Univ. of Michigan, Ann Arbor, Michigan 48109-1109*

## Abstract

It is shown analytically that every static, spherically symmetric solution to the Einstein Yang Mills equations with SU(2) gauge group that is defined in the far field has finite ADM mass. Moreover, there can be at most two horizons for such solutions. The three types of solutions possible, Bartnik–McKinnon particle-like solutions, Reissner–Nordström–like solutions, and black hole solutions having only one horizon are distinguished by the behavior of the metric coefficients at the origin.

## Introduction

The Einstein Yang Mills equations with SU(2) gauge group, derived in [BM] for static, spherically symmetric solutions in the magnetic ansatz, give a classical (non quantum mechanical) description of gravity coupled to a nuclear force modeled by a Yang Mills field. The unknowns of the equations are the metric and the connection. We may write the metric as

$$ds^2 = -A(r) B(r)^{-2} dt^2 + A(r)^{-1} dr^2 + r^2( d\theta^2 + \sin^2\theta\, d\phi^2)$$

and the Yang/Mills curvature 2–form as

$$F = w'\, \tau_1\, dr \wedge d\theta + w'\, \tau_2\, dr \wedge (\sin\theta\, d\phi) - (1 - w^2)\tau_3\, d\theta \wedge (\sin\theta\, d\phi).$$

Here $A(r) B(r)^{-2}$, $A(r)^{-1}$, and $w(r)$ denote the unknown metric and connection coefficients, respectively, prime denotes the derivative with respect to $r$, the Schwarzschild coordinate, and $\{\tau_1, \tau_2, \tau_3\}$ form a basis for $su(2)$, the Lie algebra of SU(2).

The EYM equations in this framework form a system of three ordinary differential equations:

(1.1) $\quad r A' + (1 + 2 w'^2) A = 1 - W^2/r^2$

(1.2) $\quad r^2 A w'' + r \Psi(r) w' + w W = 0$

(1.3) $\quad r B'/B = 2 w'^2$

where we set $W = 1 - w^2$ and $\Psi(r) = 1 - A - W^2/r^2$

Much effort has gone into studying this system of equations–hundreds of papers have appeared. See [GV1] for an extensive bibliography. Much of the effort has been directed towards proving the existence of solutions of various types: particle–like solutions c.f. ([BM], [KM], [SYWM], [SW1], [BFM]), black hole solutions c.f. ([B], [BP], [GV2], [GV3], [LM], [SWY]), Reissner–Nordström–like solutions c.f.[SW4], "bag of gold" solutions c.f.[BFM].

There have also been results concerning the uniqueness of solutions. For example, it is shown in [BFM] that particle–like solutions are classified by nodal class. Also, it is shown in [SW2] that any (static, spherically symmetric) solution to the EYM equations that is defined in the far field and is regular, that is, $A(r) > 0$ for $r \gg 1$, is either a particle–like solution, a black hole solution, or a Reissner–Nordström–like solution, that is, a solution that has $A(r) > 1$ for some $r$. (A solution $(A, w, B)$ is said to be defined in the far field if there is a $r_0 > 0$ such that for all $r > r_0$ the functions are defined and differentiable and equations 1.1 –1.3 are satisfied.)

In this paper we consider static, spherically symmetric solutions to the EYM equations that are defined in the far field.

Note: since equations 1.1, 1.2 do not involve B we need only discuss w and A in what follows; equation 1.3 can always be solved for B.

The first objective is to answer a question posed in [SW5]; are all solutions to equations 1.1, 1.2 that are defined in the far field regular solutions? That is, if a solution $(w, A)$ to 1.1, 1.2 is defined for $r > r_0$, is $A(r) > 0$ for $r \gg 1$? The answer, given by Theorem 9, is yes. It is shown that any solution to equations 1.1, 1.2 that is defined in the far field has finite ADM mass [ADM]; in particular, A(r) > 0 for large r. Thus, any solution defined in the far field is asymptotically flat space-time.

It was shown in [SW5] that any regular solution to the EYM equations that is defined in the far field is actually defined for all r > 0. Combining this result with theorem 9 quoted above we can say that any solution to the EYM equations that is defined in the far field is actually defined for all r > 0; we then ask about the behavior of the solution near r = 0.

If a solution is defined in the far field and A(r) > 0 for all r > 0 then the solution must be either a particle–like solution or a Reissner–Nordström– like solution with a naked singularity at the origin [SW2]. What can we say about solutions if there is a horizon, that is, if A(ρ) = 0 for some ρ? For example, can there be horizons within horizons? Are there solutions defined for all r ≥ 0 that satisfy particle like initial conditions at r = 0 and have A > 0 in the far field but for which A has 2 or more zeros? It was shown in [SW5] that the zeros of A are isolated and can only accumulate at r = 0; moreover, there are at most 2 zeros of A for r > 1. It was shown in ([Z]) that 0 is <u>not</u> an accumulation point for the zeros of A. In section 2 we give a very simple argument for a sharper result: we show that A can have at most 2 zeros. This answers another question posed in [SW5]. Moreover, if A has 2 zeros then it must be a Reissner–Nordström– like solution with a singularity at the origin (Proposition 22). There can also be solutions for which A has exactly one zero and for these solutions we show that $\liminf_{r \to 0} A(r) = -\infty$ (Theorem 20). This confirms a conjecture of [SW5].

We thus have a trichotomy: for solutions defined in the far field the behavior of the solutions at r = 0 can be of three types: a) $\lim_{r \to 0} A(r) = 1$, b) $\lim_{r \to 0} A(r) = \infty$, or c) $\liminf_{r \to 0} A(r) = -\infty$ (Theorem 23).

Solutions of type a must be Bartnik –McKinnon particle-like solutions; see Proposition 22 and Theorem 3.7 of [SW2] . Solutions of type b must be

Reissner–Nordström–like solutions; see theorem 6.3 [SW4]. The classic Schwarzschild solution, $w(r) \equiv \pm 1$, $A(r) = 1 - 2M/r$, is of type c.

Solutions of type c with nontrivial gauge field have not been shown to exist; Reissner–Nordström–like solutions with nontrivial gauge field have only been proved to exist when there is a naked singularity [SW4]. There is some numerical evidence for the existence of Reissner–Nordström–like solutions for which A has 2 zeros ([BLM], [DGZY1], [DGZY2]). See also [GV1], page 49. On the other hand, for any $\rho > 0$ there exists a countable number of solutions to the EYM equations 1.1-1.3 defined in the far field, distinguished by nodal class, and having horizon $\rho$ ([SWY], [BFM]). By the results of [SW5] such solutions are defined for all $r > 0$, have nontrivial gauge field and the metric coefficient A has one or two zeros. Thus, there must exist solutions to equations 1.1–1.3 with nontrivial gauge field of type b or type c; possibly both types of solutions occur.

The paper is organized as follows: in section 1 it is shown that for any solution to 1.1, 1.2 defined in the far field A can have at most 2 zeros. In section 2 it is shown that for any solution to 1.1, 1.2 defined in the far field A must be positive for large r. In section 3 solutions having horizons are discussed and the trichotomy theorem is presented.

I would like to thank Piotr Bizon for helpful comments.

§ 1

In this section we show that if $(w(r), A(r))$ is any solution to the EYM equations that is defined in the far field, that is, for all $r > r_0$ for some $r_0$, then $A(r)$ has at most two zeroes.

**Definition** For any solution $(w, A)$ of the EYM equations 1.1, 1.2 we set $L(r) = 1 + A(1 + 2\, w'^2) - W^2/r^2$.

**Note 0**. We can also write $L = r A' + 2A(1 + 2 w'^2)$. It follows from 1.1, 1.2 that the function L satisfies the equation $rL'(r) = L(1 - 2 w'^2) + 2 (W^2/r^2 - A)$.

**Proposition 1** If $L(r_1) = 0$ then $L'(r_1) > 0$.

Proof: If $L(r_1) = 0$ then $r_1 L'(r_1) = 2 (W^2/r_1^2 - A)$ which is clearly positive if $A < 0$. If $A \geq 0$ we use $L = 0$ to write $W^2/r_1^2 - A = 1 + 2Aw'^2 > 0$.

**Corollary 2** If $L(r_1) = 0$ then $L(r) > 0$ for $r > r_1$ and $L(r) < 0$ for $r < r_1$; hence, $L(r) = 0$ can have at most one solution.

**Corollary 3** If $A(\tau) = 0$ and $A'(\tau) < 0$ then $A(r) > 0$ for all $r < \tau$.

Proof: By Note 0, $L(\tau) = \tau A'(\tau) < 0$. If $A(r) = 0$ for some $r < \tau$, let $\sigma$ be the largest such $r < \tau$ with $A(r) = 0$. Since $A(r) > 0$ for $r$ near $\tau$, $r < \tau$, we have $A'(\sigma) \geq 0$ and hence, $L(\sigma) = \sigma A'(\sigma) \geq 0$. But by the above proposition, $L(r) < 0$ for all $r < \tau$ which is a contradiction.

**Corollary 4** If $A(\sigma) = 0$ and $A'(\sigma) > 0$ then $A(r) > 0$ for all $r > \sigma$.

Proof: By Note 0, $L(\sigma) = \sigma A'(\sigma) > 0$. If $A(r) = 0$ for some $r > \sigma$, let $\tau$ be the smallest such $r > \sigma$ with $A(r) = 0$. Since $A(r) > 0$ for $r$ near $\sigma$, $r > \sigma$, we have $A'(\tau) \leq 0$ and hence, $L(\tau) = \tau A'(\tau) \leq 0$. But by the above proposition $L(r) > 0$ for all $r > \sigma$ which is a contradiction.

**Corollary 5** If $A(\gamma) = 0 = A'(\gamma)$ then $A(r) > 0$ for all $0 < r \neq \gamma$.

Proof: We first show that if $A(\gamma) = 0 = A'(\gamma)$ then $A''(\gamma) > 0$. Differentiating equation 1.1 at $\gamma$ gives $A''(\gamma) = 2W^2/\gamma^3 + 4wWw'/\gamma^2$. It follows easily from equation 1.2 that $w(\gamma) = 0$ since $A(\gamma) = 0$, $1 - W^2/\gamma^2 = 0$, and $\gamma > 0$, thus $A''(\gamma) = 2/\gamma^3 > 0$. In particular, $A(r) > 0$ for $r$ near $\gamma$, $r \neq \gamma$.

By Note 0, $L(\gamma) = \gamma A'(\gamma) = 0$ since $A(\gamma) = 0$. If $A(r) = 0$ for some $r > \gamma$, let $\tau$ be the smallest such $r > \sigma$ with $A(r) = 0$. Since $A(r) > 0$ for $r$ near $\gamma$, $r > \gamma$, we have $A'(\tau) \leq 0$ and hence, $L(\tau) \leq 0$. But by the above proposition $L(r) > 0$ for all $r > \gamma$ which is a contradiction.

If $A(r) = 0$ for some $r < \gamma$, a symmetric argument produces a $\sigma < \gamma$ with $L(\sigma) \geq 0$, another contradiction.

**Theorem 6** If $(w, A)$ is a solution to the equations 1.1, 1.2 defined for $0 \leq r_1 < r < r_2$ then $A$ has at most two zeroes.

Proof: If $A(\gamma) = 0 = A'(\gamma)$ for some $r_1 < \gamma < r_2$ then $A(r) > 0$ for all $r \neq \gamma$, $r_1 < r < r_2$ by the above corollary and hence, $A$ has only one zero. Otherwise, if $A(\sigma) = 0$ and $A'(\sigma) > 0$ for some $r_1 < \sigma < r_2$ then by corollary 4, $A(r) > 0$ for all $r > \sigma$ so any zero of $A$ must occur for $r < \sigma$. Symmetrically, if $A(\tau) = 0$ and $A'(\tau) < 0$ for some $r_1 < \tau < \sigma$ and $\tau$ is the largest such $r$ then by corollary 3, $A(r) > 0$ for all $r < \tau$. Thus $A$ can have at most two zeroes.

**Remark 7** The theorem actually shows a bit more, namely, if $A$ has two zeroes at $\tau$ and $\sigma$ say, with $A'(\sigma) > 0$ and $A'(\tau) < 0$ then $\tau < \sigma$. Thus, for example, there do not exist local solutions to equations 1.1, 1.2 that are negative for $r$ near 0, then positive for an interval of $r$ then negative.

**Remark 8** It follows from [SW3] that there are no solutions to equations 1.1, 1.2 defined in the far field having $A(\gamma) = 0 = A'(\gamma)$ if $A(r) > 0$ for large $r$; thus, corollary 5 is not needed for the proof of theorem 6 if $r_2 = \infty$. However, we include corollary 5 because the proof is so simple and because it does not require the solution to satisfy $A(r) > 0$ for large $r$ nor is the solution assumed to exist for large $r$.

## § 2

In this section we show that any solution to equations 1.1, 1.2 that is defined in the far field has $A(r) > 0$ for large $r$. Such solutions were dubbed "regular" in [SW2]. An important corollary is that any solution to the spherically symmetric SU(2) Einstein/Yang–Mills equations defined in the far field has finite ADM mass. Lemmas 12 and 13 were announced earlier [SW5].

**Theorem 9** If (w, A) is a solution to equations 1.1, 1.2 defined in the far field then A(r) > 0 for large r.

Proof: Assume throughout this section that (w, A) is a solution to equations 1.1, 1.2 defined in the far field with A(r) < 0 for large r; we will derive a contradiction.

The proof requires a number of lemmas.

**Lemma 10** L(r) > 0 for large r; in particular, $1 - W^2/r^2 > 0$, $\Psi(r) > 0$ and $-2Aw'^2 < 1$.

Proof: We calculate $r(L + A)' = 2 - 2A - 2w'^2 L$. If L < 0 and A < 0 then $r(L + A)' \geq 2$ and hence, L + A > 0 for large r. Thus L > –A > 0 for large r. Next, since $1 + A(1 + 2w'^2) - W^2/r^2 > 0$ it follows that $1 - W^2/r^2 > -A(1 + 2w'^2) \geq 0$. Also, $\Psi(r) = 1 - W^2/r^2 - A > 1 - W^2/r^2 > 0$. The last assertion follows from $1 + 2Aw'^2 > W^2/r^2 - A \geq 0$.

**Lemma 11** $\liminf_{r \to \infty} 1 - \frac{W^2}{r^2} = 0$.

Proof: By lemma 10 we have $1 - \frac{W^2}{r^2} > 0$ for large r. Suppose $\liminf_{r \to \infty} 1 - \frac{W^2}{r^2} = 2\eta > 0$. Then for large r, $1 - W^2/r^2 > \eta$. Note that $(rA)' = 1 - W^2/r^2 - (1 + 2w'^2) A \geq 1 - W^2/r^2 \geq \eta$ so rA > 0 for some r. That is a contradiction.

**Lemma 12** The projection of the orbit (w,w') in the w–w' plane cannot remain in the second or fourth quadrant Q2 or Q4 for all $r > r_2$.

Proof: In Q2 and Q4 ww' < 0 so $w^2$ is decreasing; hence, $\liminf_{r \to \infty} 1 - \frac{W^2}{r^2} = 1$ contradicting lemma 11.

**Lemma 13** The projection of the orbit (w,w') in the w–w' plane cannot remain in the first or third quadrant Q1 or Q3 for all $r > r_2$.

Proof: In Q1 we have $r(A + Aw')' = -2Aw'^3 - A - \frac{2Aw'^2}{r} + 1 - \frac{W^2}{r^2} - \frac{wW}{r} > 1 - \frac{W^2}{r^2} - \frac{wW}{r}$.

Now $1 - W^2/r^2 \geq 0$ by lemma 10 and hence $w < \sqrt{r+1}$. In the interval $0 \leq w \leq \sqrt{r+1}$ the (abstract) function of w, $1 - \frac{W^2}{r^2} - \frac{wW}{r} > 1/4$ so

$r(A + Aw')' \geq 1 - \frac{W^2}{r^2} - \frac{wW}{r} > 1/4$ and thus A(1+w') > 0 for large r. But w' > 0 in Q1 and hence A > 0 which is a contradiction. Thus the orbit must leave Q1.

In Q3 we use $r(A - Aw')' = +2Aw'^3 - A - \frac{2Aw'^2}{r} + 1 - \frac{W^2}{r^2} + \frac{wW}{r} > 1 - \frac{W^2}{r^2} + \frac{wW}{r}$; the rest of the argument proceeds mutatis mutandis.

Lemmas 12 and 13 show that A < 0 implies that the projection of the orbit (w,w') in the w–w' plane must rotate; equation 1.2 shows that the rotation must be about (1, 0) or (–1, 0) or both. Lemma 11 shows that the size of the loops is unbounded.

**Lemma 14** $\limsup_{r \to \infty} 1 - \frac{W^2}{r^2} = 1$.

Proof: It is clearly sufficient to show that for any $r_0$ there is an $r > r_0$ with $1 - W^2/r^2 = 1$.

By lemma 11 there is an $r_1 > r_0$ such that $1 - W^2/r_1^2 < 1/2$ say, i.e., $w^2 - 1 > r/2 \gg 1$. Suppose that $(w(r_1), w'(r_1))$ is in Q1 (respectively Q3). By lemma 13 the orbit must exit to Q4 (respectively Q2). Then, by lemma 12, the orbit must leave Q4 (respectively Q2); that can only happen if $w^2 < 1$ and the orbit exits to

Q1(respectively Q3) or w = 0 and the orbit exits to Q3 (respectively Q1). In either case, there is an r with $w(r)^2 = 1$, i.e., $1 - W^2/r^2 = 1$.

Note that $(rA)' = 1 - W^2/r^2 - 2 w'^2 A \geq 1 - W^2/r^2 > 0$ by lemma 10 so $\lim_{r \to \infty} rA$ exists and $\lim_{r \to \infty} rA \leq 0$. Henceforth, we assume there is an $M > 0$ and $r_0$ such that $L(r) > 0$, $rA(r) > -M$ for $r > r_0$.

To prove theorem 9 we will show that $\int_{r_0}^{r_1} 1 - \frac{W^2}{r^2} dr > M$ for some $r_1$ and hence,

$$r_1 A(r_1) > r_0 A(r_0) + \int_{r_0}^{r_1} 1 - \frac{W^2}{r^2} dr > -M + M = 0.$$

Using lemmas 10 and 14 we note that there are sequences $n < x_n < y_n < z_n$ such that $1 - W^2/x_n^2 = 1$, $1 - W^2/y_n^2 = 2/3$, $1 - W^2/z_n^2 = 1/3$ and $w'(r) \neq 0$ for $x_n < r < z_n$. Note that $w(x_n)^2 = 1$, $w(y_n)^2 = 1 + \sqrt{\frac{1}{3}} y_n \approx \sqrt{\frac{1}{3}} y_n$ $w(z_n)^2 = 1 + \sqrt{\frac{2}{3}} z_n \approx \sqrt{\frac{2}{3}} z_n$.

In particular, $w(r) \leq \sqrt{r}$ for $x_n < r < z_n$ and $w(z_n) - w(y_n) > \sqrt[4]{\frac{2}{3}} \sqrt{z_n} - \sqrt[4]{\frac{1}{3}} \sqrt{y_n} \approx .144 \sqrt{z_n}$.

Now $\int_{r_0}^{\infty} 1 - \frac{W^2}{r^2} dr > \sum_{n=1}^{\infty} \int_{x_n}^{z_n} 1 - \frac{W^2}{r^2} dr$ since $1 - W^2/r^2 > 0$ by lemma 10 so to finish the proof of theorem 9 it is clearly sufficient to show $\int_{x_n}^{z_n} 1 - \frac{W^2}{r^2} dr$ is uniformly bounded away from 0, that is $\int_{x_n}^{z_n} 1 - \frac{W^2}{r^2} dr > \eta > 0$ where $\eta$ is independent of n.

Now, for $x_n < r < z_n$, we have $1 - W^2/r^2 \geq 1/3$ so $\int_{x_n}^{z_n} 1 - \frac{W^2}{r^2} dr > \int_{x_n}^{z_n} \frac{1}{3} dr = \frac{z_n - x_n}{3}$. Thus, it is sufficient to show $z_n - x_n$ is uniformly bounded away from 0. Clearly, if $z_n - x_n > 1$ for all n we are done so assume $z_n - x_n < 1$. Assume also for definiteness that $w'(r) > 0$ for $x_n < r < z_n$; the argument is similar if $w'(r) < 0$. Now $w(x_n) = 1$, $w(y_n) \approx \sqrt{\frac{y_n}{\sqrt{3}}}$, so $w(y_n) - w(x_n) \approx \frac{\sqrt{y_n}}{\sqrt[4]{3}}$ Hence, $z_n - x_n > y_n - x_n = \frac{w(y_n) - w(x_n)}{w'(\zeta)'} \approx \frac{\sqrt{y_n}}{3w'(\zeta)}$ for some intermediate $\zeta$, $x_n < \zeta < y_n$. We now complete the proof of theorem 9 by showing that $\frac{w'(r)}{\sqrt{r}}$ is bounded for $x_n \leq r \leq y_n$. Since $z_n - x_n < 1$ and $x_n \gg 1$ it follows that $\frac{w'(\zeta)}{\sqrt{y_n}}$ is bounded for $x_n \leq r \leq y_n$.

**Lemma 15** If $w'(a)^2 \geq 625\, a$ for some $a < z_n$ and $x_n > 2M$ then $w'(r)^2 \geq 625\, r$ for all $a \leq r \leq z_n$.

Proof: Let $h(r) = w'(r)^2 - 625\, r$; then $h(a) \geq 0$. We show that $h(r) = 0$ implies $h'(r) > 0$ and thus, $h$ can never become negative. Since $\Psi(r) \geq 1/3$ and $w(r) \leq \sqrt{r}$ for $a \leq r \leq z_n$ we have from equation 1.2, $h'(r)|_{h(r)=0} = 2\, w'(r)w''(r) - 625 = 2\, w'(r)\{r\Psi(r)w' + w\, W\}/(-r^2 A) - 625 \geq 2\, w'(r) \{r\, w'\, /3 - r^{3/2}\}/(-r^2 A) - 625 \geq$
$50\, r^{1/2} \{8\, r^{3/2} - r^{3/2}\}/(-r^2 A) - 625 \geq 350\, r/M - 625$ where we have used $-W < r$, $w^2 < r$, $-rA < M$. Thus, $h'(r)|_{h(r)=0} > 0$ if $r > 2\, M$.

**Lemma 16** $w'(r)^2 < 625\, r$ for $x_n < r < y_n$.

Proof: Suppose $w'(a)^2 \geq 625\, a$ for some $a < z_n$, then by lemma 15 $w'(r)^2 > 625\, r$ for all $a \leq r \leq z_n$. We now apply the estimate on $w'$ to equation 1.2. First, $r\Psi(r)w' > 25\, r^{3/2}/3 > -2\, w\, W$. Hence, $-r^2 A\, w'' = (r - r\, A - W^2/r)\, w' + w\, W \geq r\Psi(r)\, w'/2 \geq$
$rw'/6$ for all $a \leq r \leq z_n$. Thus, by lemma 10, $\frac{w''}{w'^2} \geq \frac{rw'}{3r^2(-2Aw'^2)} \geq \frac{w'}{3r}$. Integrating the left hand side of the inequality from $a$ to $z_n$ yields
$$\int_a^{z_n} \frac{w''}{w'^2} dr = \frac{-1}{w'(z_n)} + \frac{1}{w'(a)} \leq \frac{1}{w'(a)}.$$
Integrating the right hand side of the inequality from $a$ to $z_n$ yields
$$\int_a^{z_n} \frac{w'}{3r} dr > \int_{y_n}^{z_n} \frac{w'}{3r} dr \approx \int_{y_n}^{z_n} \frac{w'}{3z_n} dr = \frac{w(z_n) - w(y_n)}{3z_n} \approx \frac{\sqrt[4]{\frac{2}{3}}\sqrt{z_n} - \sqrt[4]{\frac{1}{3}}\sqrt{y_n}}{3z_n} \approx \frac{.048}{\sqrt{z_n}}. \text{ Thus,}$$
$\frac{1}{w'(a)} > \frac{.048}{\sqrt{z_n}}$ or $w'(a)^2 < 435\, z_n \approx 435\, a$ contradicting the assumption that $w'(a)^2 \geq 625\, a$.

The proof of theorem 9 is now complete.

**Corollary 17** Any solution to the equations 1.1, 1.2 that is defined in the far field is defined for all r > 0. The ADM mass of the solution is finite and A(r) ≈ 1− 2μ/r for r >> 1 where μ is the ADM mass. Moreover, either the gauge field w ≡ 0 or $\lim_{r \to \infty} w^2 = 1$.

Proof: Since the solution is regular (A(r) > 0 for r >> 1) by Theorem 9 we may invoke the results of [SW2].

## § 3

In this section we first consider solutions of 1.1, 1.2 defined in the far field with A(r) having exactly one zero. Since we know by Theorem 9 that A(r) > 0 for r >>1, we need only consider solutions defined for all r > 0 and such that A(ρ) = 0, A'(ρ) > 0 for some ρ > 0. Thus A(r) < 0 for r < ρ.

By Note 0, L( ρ ) = ρ A'(ρ) > 0.

**Proposition 18** There is a b < ρ such that L(r) < 0 for all r < b.

Proof: We will first assume L(r) > 0 for all r ≤ ρ and derive a contradiction. We have r A' = L − 2A(1 + w'$^2$) > −2A and hence, (r$^2$ A)' > 0. Hence, if $r_1 < r_2 < \rho$, $A(r_1) < \frac{r_2^2}{r_1^2} A(r_2)$ so $\lim_{r_1 \to 0+} A(r_1) = -\infty$. Since L(r) < 1 + A we have a contradiction. Thus, L(b) = 0 for some b < ρ and L(r) < 0 for all r < b by Proposition 1.

**Proposition 19** $\lim_{r \to 0+} L(r) = -\infty$.

Proof: We calculate r(L + A )' = 2 − 2A − 2 w'$^2$ L. If L < 0 and A < 0 then r(L + A )' ≥ 2 and hence, by integrating, $\lim_{r \to 0+} L + A = -\infty$. Since L < 1 + A we have $2L < 1 + L + A$ and hence the result.

**Theorem 20**. If (w, A) is a solution to equations 1.1, 1.2 defined in the far field and for which A has exactly one zero then $\liminf_{r \to 0+} A(r) = -\infty$.

Proof: We assume $0 > A(r) > M$ for $r$ near $0$ and derive a contradiction. First note that $r A'(r) < -1$ for all $r < r_1$ is not possible; a simple integration shows that $A(r) > 0$ for some $r < r_1$. Similarly, $r A'(r) > 1$ for all $r < r_2$ is not possible; again, a simple integration shows that $\lim_{r \to 0+} A(r) = -\infty$. Thus, given any $r_1 > 0$ we must have $|r A'(r)| < 1$ for some $r < r_1$.

We complete the proof with the following lemma.
**Lemma 21**. If $1 > r A'(r) > -1$, $A(r) > M$ and $r \ll 1$ then $r^2 A''(r) < -2$.

Proof of theorem using lemma 21. The lemma shows that if $-1 < r_1 A'(r_1) < 1$ for some $r_1 > 0$ then, by integrating the inequality $r^2 A''(r) < -2$, there is an $r_2 < r_1$ such that $r_2 A'(r_2) > 1$. Moreover, if $r A'(r) = 1$ for some $r < r_2$ then $r(rA'(r))'$ $= r^2 A''(r) + r A'(r) < -2 + 1 < 0$ so $r A'(r) > 1$ for all $r < r_2$. As noted above, $r A'(r) > 1$ for all $r < r_2$ is not possible so that completes the contradiction and hence, the proof of the theorem.

Proof of lemma 21: By Proposition 19, $-L \gg 1$ for $r \ll 0$; so $-2 Aw'^2 + W^2/r^2 \gg 1$ since $1 + A$ is bounded. The assumption that $|r A'(r)| < 1$ says that $-2 Aw'^2 \approx W^2/r^2$, hence, $-2 Aw'^2 \gg 1$ and $W^2/r^2 \gg 1$. Also, since $A$ is bounded and $-2 Aw'^2 \gg 1$ we see that $w'^2 \gg 1$. We then compute

$$r^2 A''(r) = 4Aw'^4 + (2 + 2A - 2\frac{W^2}{r^2})w'^2 + 8w\frac{W}{r}w' + 2A - 2 + 4\frac{W^2}{r^2}.$$ We have

$-\frac{W^2}{r^2}w'^2 + 4\frac{W^2}{r^2} \ll 0$ since $w'^2 \gg 1$. Also, $(2 + 2A - \frac{W^2}{r^2})w'^2 + 8w\frac{W}{r}w' \ll 0$ since $W^2/r^2 \gg |wW/r|$ and $w'^2 \gg |w'|$. The remaining terms are all negative and hence, $r^2 A''(r) < -2$.

Note: we are not able to prove that limit $A = -\infty$ reflecting the fact that $A$ oscillates near $r = 0$, c.f.([DGZY1], [DGZY2], [BLM], [Z]).

We now examine the behavior of solutions $(w, A)$ of equations 1.1, 1.2 defined in the far field for which $A$ has two zeros at $\tau$ and $\sigma$ say, with $A'(\sigma) > 0$ and $A'(\tau) < 0$.

**Proposition 22** A solution (w, A) to equations 1.1, 1.2 defined in the far field for which A has two zeros is a Reissner–Nordström– like solution.

Proof: By Remark 7, $\tau < \sigma$ and hence $A(r) > 0$ for r near 0. It was shown in [SW2} that any solution to 1.1, 1.2 that is defined in a neighborhood of r = 0 and that has A(r) > 0 near r = 0 is either a Reissner–Nordström– like solution or a Bartnik-McKinnon particle-like solution. Thus, it is sufficient to show that (w, A) is not a Bartnikon. As observed in Corollary 3, $L(\tau) < 0$ and hence $L(r) < 0$ for all $r < \tau$ by lemma 10. But a Bartnik-McKinnon particle-like solution has L(0) = 2 since A(0) = 1, w'(0) = 0, $w^2(0) = 1$. Thus, the solution cannot be a Bartnik-McKinnon particle-like solution and must be a Reissner–Nordström–like solution.

Note that the singularity at r = 0 is inside the horizon at $r = \rho > \tau > 0$.

We now have the Trichotomy Theorem:

**Theorem 23** Any solution of equations 1.1, 1.2 defined in the far field is defined for all r > 0. Moreover, either $\lim_{r \to 0+} A(r) = 1$, $\lim_{r \to 0+} A(r) = +\infty$, or $\liminf_{r \to 0+} A(r) = -\infty$.

Proof: If A(r) > 0 for all r > 0 then by the results of [SW2] mentioned above the solution is either a Bartnik-McKinnon particle-like solution for which $\lim_{r \to 0+} A(r) = 1$ or a Reissner–Nordström– like solution for which $\lim_{r \to 0+} A(r) = +\infty$ [SW4]. If A has one zero then by Theorem 20 $\liminf_{r \to 0+} A(r) = -\infty$. Finally, if A has two zeros then the solution is a Reissner–Nordström– like solution by Proposition 22 and hence, $\lim_{r \to 0+} A(r) = +\infty$. By theorems 6 and 9 there are no other cases to consider.